\documentclass[twocolumn,notitlepage,]{article}
\def\1{{\bf 1}}

\def\Tr{{\rm Tr}}
\def\C{{\bf C}}
\def\R{{\bf R}}
\def\E{{\cal E}}

\def\mfr#1/#2{\hbox{${{#1} \over {#2}}$}}
\def\const.{{\rm const.}}
\def\beq{\begin{equation}}
\def\eeq{\end{equation}}
\def\beqa{\begin{eqnarray}}
\def\eeqa{\end{eqnarray}}

\begin{document}
\setlength{\unitlength}{1.0cm}

\def\boxit#1{\thinspace\hbox{\vrule\vtop{\vbox{\hrule\kern1pt
\hbox{\vphantom{\tt/}\thinspace{\tt#1}\thinspace}}\kern1pt\hrule}\vrule}
\thinspace}

\font\eightit=cmti8
\def\ve{{\varepsilon}}
\def\C{{\bf C}}
\def\D{{\rm D}}
\def\E{{\rm E}}
\def\R{{\bf R}}
\def\Z{{\bf Z}}
\def\L{{\Lambda}}
\def\l{{\lambda}}
\def\d{{\rm d}}
\def\Q{{{\rm C}_{\Lambda}}}
\def\X{{\underline{X}}}
\def\Tr{{\rm Tr}}


{\bf THOMAS-FERMI THEORY} -- Sometimes called the `statistical theory',
it was invented by L. H. Thomas\cite{TH} and E. Fermi\cite{EF}, shortly
after Schr\"odinger invented his quantum-mechanical wave equation, in
order to approximately 
describe the \emph{electron density}, $\rho(x)$, $x\in \R^3$, and the
\emph{ground state energy}, $E(N)$ for a large atom or molecule 
with a large number,
$N$, of electrons. Schr\"odinger's equation, which would give the 
exact density and energy, cannot be easily handled when $N$ is large.

A starting point for the theory is the \emph{TF energy functional}.
For a molecule with $K$ nuclei of charges $Z_i >0$ and locations
$ R_i\in \R^3 \ (i=1,...,K)$, it is

\begin{eqnarray}
{\cal E}(\rho) & := & \frac{3}{5} \gamma \int_{\R^3}\rho(x)^{5/3} \ \d x -
            \int_{\R^3} V (x) \rho (x) \ \d x \nonumber \\ 
            & + & \frac{1}{2} \int_{\R^3} \int_{\R^3} ~ \frac{\rho(x)
            \rho (y)}{|x-y|} \ \d x \d y + U
\end{eqnarray}
in suitable units. Here,
\begin{eqnarray*}
V(x) &=& \sum_{j=1}^{K} Z_j |x-r_j|^{-1}\ ,\\
U &=& \sum_{1 \leq i < j
\leq K} Z_i Z_j |R_i - R_j|^{-1} \ ,
\end{eqnarray*}
and $\gamma = (3 \pi^2)^{2/3}$. The constraint on $\rho$ is $\rho (x)
\geq 0$ and $\int_{\R^3} \rho = N$. The functional $\rho \rightarrow
{\cal E}(\rho)$ is convex.\\

\noindent The justification for this functional is this:

\noindent $\bullet$ The first term is roughly the minimum
quantum-mechanical kinetic 
energy of $N$ electrons needed to produce an electron density $\rho$.\\
\noindent $\bullet$ The second term is the attractive interaction of
the $N$ electrons with the $K$
nuclei, via the {\it Coulomb potential} $V$.\\
\noindent $\bullet$ The third is approximately the electron-electron
repulsive energy.\\
\noindent $\bullet$ $U$ is the nuclear-nuclear repulsion and is an
important constant.\\

The {\it TF energy} is defined to be
\[
E^{\rm{TF}}(N) = \inf \{ {\cal E}(\rho) : \rho \in
L^{5/3}, \int \rho = N, \rho \geq 0 \} \ ,
\]
i.e., the TF energy and density is obtained by minimizing ${\cal E}(\rho)$ with
$\rho \in L^{5/3} (\R^3)$ and $\int \rho = N$. The {\it Euler-Lagrange
equation}, called the {\it Thomas-Fermi equation}, is
\begin{equation}
\gamma \rho (x)^{2/3} = \left[ \Phi (x) - \mu \right]_+,
\end{equation}
where $[a]_+$ = $\max \{0,a\}$, $\mu$ is some constant ({\it Lagrange
multiplier}) and $\Phi$ is the {\it TF potential}:
\begin{equation}
\Phi (x) = V(x) - \int_{\R^3} |x-y|^{-1} \rho(y) \ \d y.
\end{equation}
The following essential mathematical facts about the TF equation were
established by 
E.H. Lieb and B. Simon~\cite{LS} ({\it cf.} the review
article~\cite{EL}).

\begin{enumerate}
\item There is a density $\rho^{\rm{TF}}_{N}$ that minimizes ${\cal
E}(\rho)$
if and 
only if $N \leq Z : = \sum_{j=1}^{K} Z_j$. This $\rho^{\rm{TF}}_{N}$ is
unique and it satisfies the TF equation (2) for some $\mu \geq
0$. Every positive solution, $\rho$, of (2) is a minimizer of (1) for $N = \int
\rho$. If $N >  Z$ then $E^{\rm{TF}}(N) = E^{\rm{TF}}(Z)$ and any
minimizing sequence converges weakly in $L^{5/3}(\R^3)$ to
$\rho^{\rm{TF}}_Z$.
\item $\Phi (x) \geq 0$ for all $x$. (This need not be so for the
real Schr\"odinger $\rho$.)
\item $\mu = \mu(N)$ is a strictly monotonically decreasing function
of $N$ and $\mu(Z)=0$ (the {\it neutral case}). $\mu$ is the {\it
chemical potential}, namely  
\[
\mu (N) = -\frac{\partial
E^{\rm{TF}}(N)}{\partial N} \ .
\]
$E^{\rm{TF}}(N)$ is a strictly convex, decreasing function of $N$ for
$N \leq Z$ and $E^{\rm{TF}}(N) = 
E^{\rm{TF}}(Z)$ for $N \geq Z$. If $N < Z$, $\rho^{\rm{TF}}_N$ has compact
support.
\end{enumerate}

When $N=Z$, (2) becomes $\gamma \rho^{2/3} = \Phi$. By applying the
Laplacian $\Delta$ to both sides we obtain
$$ 
- \Delta \Phi (x) + \ 4
\pi \gamma^{-3/2} \Phi (x)^{3/2} =4\pi  \sum_{j=1}^{K} \ Z_j \
\delta (x-R_j) \ , 
$$ which is the form in which the TF equation is usually
stated (but it is valid only for $N=Z$).

An important property of the solution is {\it Teller's
theorem}~\cite{ET} (proved rigorously in \cite{LS}) which implies that
the TF molecule is always unstable, i.e., for each $N \leq Z$ there
are $K$ numbers $N_j \in (0,Z_j)$ with $\sum_j N_j = N$ such that 
\begin{equation}
E^{\rm{TF}}(N) >  \sum^{K}_{j=1} \ E^{\rm{TF}}_{\rm{atom}} (N_j, Z_j) \ ,
\end{equation}
where $E_{\rm{atom}}^{\rm{TF}} (N_j, Z_j)$ is the TF energy with $K=1,
Z=Z_j$ and $N=N_j$. The presence of $U$ in (1) is crucial for this
result. The inequality is strict. Not only does $E^{\rm{TF}}$ decrease when
the nuclei are pulled infinitely far apart (which is what (4) says)
but any dilation of the nuclear coordinates $(R_j \rightarrow \ell R_j,
\ell >  1)$ will decrease $E^{\rm{TF}}$ in the neutral case ({\it
positivity of the pressure})~\cite{EL},~\cite{BL}. This theorem plays an
important role in the {\it stability of matter}.

An important question concerns the connection between $E^{\rm{TF}}(N)$ and
$E^{\rm{Q}}(N)$, the ground state energy (= infimum of the spectrum) of
the Schr\"odinger operator, $H$, it was meant to approximate.
\[
H = - \sum^{N}_{i=1} \left[ \Delta_i + V(x_i) \right] + \sum_{1 \leq i
< j \leq N} |x_i - x_j|^{-1} + U \ ,
\]
which acts on the antisymmetric functions $\wedge^{N} L^2 (\R^3;
\C^2)$ (i.e., functions of space and spin). It used to be believed
that $E^{\rm{TF}}$ is asymptotically exact as $N \rightarrow \infty$ but
this is not quite right; $Z \rightarrow \infty$ is also needed. Lieb
and Simon~\cite{LS} proved that if we fix $K$ and $Z_j/Z$ and we set
$R_j = Z^{-1/3} R_j^0$, with fixed $R_j^0 \in \R^3$, and set $N = \lambda
Z$, with $0 \leq \lambda < 1$ then
\begin{equation}
\lim_{Z \rightarrow \infty} \ \ E^{\rm{TF}}(\lambda
Z)/E^{\rm{Q}}(\lambda Z) = 1 \ .
\end{equation}
In particular, a simple change of variables shows that
$E^{\rm{TF}}_{\rm{atom}} (\lambda, Z) = Z^{7/3} E^{\rm{TF}}_{\rm{atom}} (\lambda, 1)$
and hence the true energy of a large atom is asymptotically proportional
to $Z^{7/3}$. Likewise, there is a well-defined sense in which the
quantum mechanical density converges to $\rho^{\rm{TF}}_N$ ({\it
cf.}~\cite{LS}).

The TF density for an atom located at $R=0$, which is spherically
symmetric, scales as
\begin{eqnarray*}
\rho^{\rm{TF}}_{\rm{atom}} (x; N = \lambda Z, Z) &=&\\
Z^2\rho^{\rm{TF}}_{\rm{atom}} (&Z^{1/3} x;& N =
\lambda, Z = 1) \ .
\end{eqnarray*}
Thus, a large atom (i.e., large $Z$) is {\it smaller} than a $Z = 1$
atom by a factor $Z^{-1/3}$ in radius. Despite this seeming paradox,
TF theory gives the correct electron density in a real atom --- so far
as the bulk of the electrons is concerned --- as $Z \rightarrow
\infty$

Another important fact is the large  $|x|$ asymptotics of
$\rho^{\rm{TF}}_{\rm{atom}}$ for a neutral atom. As $|x| \rightarrow
\infty$,
$$
\rho^{\rm{TF}}_{\rm{atom}} (x, N = Z, Z) \sim \gamma^3 (3/\pi)^3 |x|^{-6} \ ,
$$
{\it independent} of $Z$. Again, this behavior agrees with quantum
mechanics --- on a length scale $Z^{-1/3}$, which is where the bulk of
the electrons are to be found.

In light of the limit theorem (5), Teller's theorem can be understood
as saying that as $Z \rightarrow \infty$ the quantum mechanical binding
energy of a molecule is of lower order in $Z$ than the total ground
state energy.  Thus, Teller's theorem is not a defect of TF
theory (although it is sometimes interpreted that way) but an important
statement about the true quantum mechanical situation.

For finite $Z$ one can show, using the  Lieb-Thirring
inequality~\cite{LT} and the Lieb-Oxford inequality~\cite{LO}, that
$E^{\rm{TF}}(N)$, with a modified $\gamma$, gives a lower bound to
$E^{\rm{Q}}(N)$. 

Several `improvements' to Thomas-Fermi theory have been proposed, but
none have a fundamental significance in the sense of being `exact'
in the $Z \rightarrow \infty$ limit. The von Weizs\"acker correction
consists in adding a term
\[
(\rm{const.}) \int_{\R^3} | \nabla \sqrt{\rho (x)} |^2 \ \d x
\]
to ${\cal E} (\rho)$. This preserves the convexity of ${\cal E}(\rho)$
and adds (const.)$Z^2$ to $E^{\rm{TF}}(N)$ when $Z$ is large. It also has the
effect that the range of $N$ for which there is a minimizing $\rho$ is
extend from [0,Z] to [0,Z + (const.) K].

Another correction, the  {\it Dirac exchange energy}, is to add
\[
-({\rm{const.}}) \int_{\R^3} \rho (x)^{4/3} \ \d x
\]
to ${\cal E}(\rho)$. This spoils the convexity but not the range [0,Z]
for which a minimizing $\rho$ exists {\it cf.}~\cite{LS} for both of
these corrections.

When a uniform external magnetic field $B$ is present, the
operator $- \Delta$ in $H$ is replaced by 
$$
|i\nabla + A (x)|^2 +
\sigma \cdot B(x) \ ,
$$
with curl $A=B$ and $\sigma$ denoting the Pauli spin matrices. This
leads to a modified TF theory that is asymptotically exact as $Z
\rightarrow \infty$, but the theory depends on the manner in which $B$
varies with $Z$. There are five distinct regimes and theories: $B \ll
Z^{4/3}, B \sim Z^{4/3}, Z^{4/3} \ll B \ll Z^3, B \sim Z^3, \gg Z^3$.
These theories~\cite{LSY1},~\cite{LSY2} are relevant for neutron stars.
Another class of TF theories with magnetic fields is relevant for
electrons confined to two-dimensional geometries (quantum
dots)~\cite{LSY3}. In this case there are three regimes. A convenient
review is~\cite{LSY4}.

Still another modification of TF theory is its extension from a
theory of the ground states of atoms and molecules (which corresponds to zero
temperature) to a theory of positive temperature states of large
systems such as stars (cf.~\cite{JM},~\cite{WT}).

\bigskip

\rightline{\it Elliott H. Lieb}
\rightline{\it Departments of Mathematics and Physics}
\rightline{\it Princeton University}

\bigskip
\footnoterule
\bigskip
\noindent {\copyright 1998 by Elliott H. Lieb}


\begin{thebibliography}{99}

\bibitem[BL]{BL} BENGURIA, R. AND LIEB, E.H.: `The positivity of the
pressure in Thomas-Fermi theory', {\it Commun. Math. Phys.} {\bf 63}
(1978), 193-218.  {\it Errata} {\bf 71}, (1980), 94.

\bibitem[EF]{EF} FERMI, E.: `Un metodo statistico per la
determinazione di alcune priorieta dell'atome', {\it Rend. Accad. Naz. Lincei}
{\bf 6} (1927), 602-607.

\bibitem[EL]{EL} LIEB, E.H.: `Thomas-Fermi and related theories of
atoms and molecules', 
{\it Rev. Mod. Phys.} {\bf
53} (1981), 603-641.  Errata {\bf 54} (1982), 311.


\bibitem[LO]{LO}  LIEB, E.H. AND OXFORD, S.: `An improved lower bound
on the indirect coulomb energy', {\it Int. J. Quant. Chem.} {\bf 19}
(1981), 427-439.

\bibitem[LS]{LS} LIEB, E.H. AND SIMON, B.: `The Thomas-Fermi theory of
atoms, molecules and solids', {\it Adv. in Math} {\bf 23} (1977), 22-116.

\bibitem[LSY1]{LSY1} LIEB, E.H., SOLOVEJ, J.P., AND YNGVASON, J.:
`Asymptotics of heavy atoms in high magnetic fields:  I. lowest Landau
band region', {\it Commun. Pure Appl. Math.} {\bf 47} (1994), 513-591.
 
\bibitem[LSY2]{LSY2} LIEB, E.H., SOLOVEJ, J.P., AND YNGVASON, J.:
`Asymptotics of heavy atoms in high magnetic fields:  II.
semiclassical regions', {\it Commun. Math. Phys.} {\bf 161} (1994), 77-124.

\bibitem[LSY3]{LSY3}  LIEB, E.H., SOLOVEJ, J.P., AND YNGVASON, J.:  
`Ground states of large quantum dots in magnetic fields', {\it
Phys. Rev. B} {\bf 51} (1995) 10646-10665.

\bibitem[LSY4]{LSY4} LIEB, E.H., SOLOVEJ, J.P., AND YNGVASON, J.: 
`Asymptotics of natural and artificial atoms in strong magnetic
fields, in W. THIRRING (ed.): {\it The stability of matter: from atoms to
stars, selecta of E. H. Lieb}, second edition, Springer, 1997, 
pp. 145-167.

\bibitem[JM]{JM} MESSER, J.: `Temperature dependent Thomas-Fermi
theory': Vol. 147 of {\it Lecture Notes in Physics}, Springer, 1981.

\bibitem[ET]{ET} TELLER, E.,: `On the stability of molecules in
Thomas-Fermi theory', {\it Rev. Mod. Phys.} {\bf 34} (1962), 627-631.

\bibitem[LT]{LT} LIEB, E.H. AND THIRRING W.: `Inequalities for the
moments of the eigenvalues of the Schr\"odinger Hamiltonian and their
relation to Sobolev inequalities', in E. LIEB, B. SIMON, A. WIGHTMAN
(eds.): {\it `Studies in Mathematical
Physics}', Princeton University Press, 1976, pp. 269-303.

\bibitem[WT]{WT} THIRRING, W.: `A course in mathematical physics':
Vol. 4, Springer, 1983, pp. 209-277.

\bibitem[TH]{TH} THOMAS, L.H.: `The calculation of atomic fields',
{\it Proc. Camb. Phil. Soc.} {\bf 23} (1927), 542-548.

\end{thebibliography}
\end{document}